# Evaluating the Applicability of Bandwidth Allocation Models for EON Slot Allocation

Rafael F. Reale[1], Romildo M. S. Bezerra[1], Gilvan Durães[2], Alexandre C. Fontinele[3], André C. B. Soares[3] and Joberto S. B. Martins[4]

[1]Federal Institute of Bahia (IFBA) -  reale@ifba.edu.br
[2]Baiano Federal Institute (IFBaiano) - gilvan.duraes@catu.ifbaiano.edu.br
[3]Federal University of Piauí (UFPI) - alexandrefontinele@gmail.com, andre.soares@ufpi.edu.br
[4]Salvador University (UNIFACS) - joberto.martins@unifacs.br

*Abstract*—Bandwidth Allocation Models (BAMs) configure and handle resource allocation (bandwidth, LSPs, fiber, slots) in networks in general (IP/MPLS/DS-TE, optical domain, other). In this paper, BAMs are considered for elastic optical networks slot allocation targeting an improvement in resource utilization. The paper focuses initially on proposing a BAM basic configuration parameter mapping suitable for elastic optical circuits. Following that, MAM, RDM and ATCS BAMs are applied for elastic optical networks resource allocation and the overall network resource utilization is evaluated. A set of simulation results and BAM "behaviors" are presented as a proof of concept to evaluate BAM´s applicability for elastic optical network slot allocation. Authors argue that a slot allocation model for EON based on BAMs may improve utilization by dynamically managing the aggregated traffic profile.

## I. Introduction and Motivation

Nowadays, there is a growing interest in investigating the optical network architecture without the fixed wavelength grid (named *gridless*), in which network elements will support flexible bandwidth lightpaths. Thus, an optical path can occupy a free band of the spectrum exactly in accordance with the client's traffic demand. These networks were introduced in [1] and are known in the literature as Spectrum-Sliced Elastic Optical Path Network or Elastic Optical Networks [1][2] (referred to as EON hereafter).

Bandwidth Allocation Models (BAMs) [3] [4] [5] can provide a new resource (slice) allocation strategy for EON. In effect, BAMs will provide EON network operations and management with a set of allocation strategies (BAM models and applications traffic class mappings and configuration) that may be dynamically adapted to current users demands in terms of application characteristics and traffic volume (the dynamic behavior of network clients and users).

A first BAM contribution to allocate resources in EON is its ability to map services, users, applications requirements and priorities on a set of classes (TC – Traffic Classes). BAMs allow the configuration and management of application groups (TCs) that will have assigned resources and will be treated in an equivalent way in terms of resource allocation by the bandwidth allocation model used. The applications mapping considers a Service Level Agreement – SLA to comply with the QoS (Quality of Service) parameters that must be ensured by the service provider over the optical network.

Beyond that, BAM deployment will dynamically allow network management to adapt the resource allocation strategy in relation to the dynamicity of input traffic. In effect, each BAM model applied results in distinct network resource allocation strategies and distinct network performance. As an example, in case the network is submitted to a burst of high throughput consuming traffic like video streams, BAM utilization allows model switching in such a way that network utilization could be improved.

This paper proposes to evaluate the applicability of BAMs for EON resource allocation focusing on its ability to configure network users grouping and to adapt dynamically in relation to the type and volume of input traffic. In terms of this preliminary evaluation, it is expected that BAM's adoption will result in improving resources utilization for EON. The models proposed, as such, do not focus on service fairness and this aspect is not considered in this initial evaluation.

The paper is organized as follows. Section II presents the related work; Section III presents a brief review and analysis of the most applied and referenced bandwidth allocation models (MAM, RDM, and ATCS - AllocTC-Sharing) that are considered for EON. Section IV proposes a mapping of bandwidth allocation model's characteristics for EON. Simulation results are presented in Section V as a proof of concept for evaluating the benefits of BAMs applicability in EON. Finally, the conclusions are presented in Section VI.

## II. Related Work

The literature presents studies that propose and evaluate strategies for EON resource allocation considering different scenarios and adopting distinct classes of service in [6] [7] [8][9][10][11].

In [6] is presented a model and simulation of partitioning and sharing spectrum based strategies in EON for provisioning of different service demands. BAM approach proposed adopts a distinct strategy in which there is no reservation and, in effect, traffic demand will compete for resources inside each specific class

In [7], Callegati et. al. propose and analyze a trunk reservation based strategy that aims to reduce the spectrum fragmentation. This approach focused on spectrum fragmentation and BAM-based resource allocation focus on network utilization among BAM models.

In [8], Hesselbach et. al. propose a specific resource allocation method for EON based on a modified RDM. The authors evaluate the mechanism proposed under different priority classes in a simple example. The result shows the better performance of the proposal when compared with MAM and RDM in terms of link utilization and acceptance ratio. This research adopts and compares only MAM and RDM modified models. Our proposal goes beyond and evaluates MAM, RDM and AllocTC-Sharing demonstrating that AllocTC-Sharing behaves better than MAM and RDM approaches and, as such, presents an incremental contribution in terms of BAMs.

Authors in [9] use the RDM to allocate bandwidth for intra-Optical Network Unit in an Ethernet Passive Optical Network. The proposal achieves a superior performance when is compared with other two dynamic hierarchical bandwidth allocation algorithms in terms of bandwidth utilization, packet delay and fairness. This is another example where BAMs can be used to allocate resources. The comparison with the proposed EON approach is not direct but the reference illustrates another way BAMs can be adopted in the optical domain.

Authors in [10] propose and evaluate the operation of RDM model with four class types in WDM networks in terms of blocking probability and it's another example of approach in applying BAMs in optical domain. Our proposal also evaluates blocking probability, but focus on utilization using an additional model (ATCS) with a different number and distinct traffic classes configurations. Both papers are aligned with our approach providing an incremental evaluation in terms of ATCS model.

### III. BANDWIDTH ALLOCATION MODELS

In IP/MPLS/DS-TE networks, BAMs are used to define rules and limits for link allocation by defining Bandwidth Constraints (BCs) for traffic classes (TCs) [3][4]. In practice, these models effectively define how bandwidth resources are obtained and shared among applications and/or clients.

The main proposed BAMs for IP/MPLS/DS-TE networks like MAM – Maximum Allocation Model, RDM – Russian Dolls Model and ATCS – AllocTC-Sharing have distinct operational characteristics, from now on called "behaviors". Each BAM treat the input traffic profile with a different "behavior" and, as such, optimization may differ among distinct BAM models.

Table 1 resumes the expected BAM´s resource (link, lambda, slot) allocation "behavior" characteristics for distinct network traffic profiles. It is observed that utilization might be compromised depending on the input traffic profile and currently adopted BAM [12].

The MAM model targets network traffic profiles in which a strong isolation between traffic classes (TCs) is required [3]. This characteristic is valid for both optical and IP traffic demands. In this model, TCs use only private resources and there is no resource sharing among TCs (Table 1).

TABLE I. BAM "BEHAVIORS" AND OTHER OPERATIONAL CHARACTERISTICS

| BAM – General Behavioral Characteristics | MAM | RDM | ATCS |
|---|---|---|---|
| Resource _utilization_ with a traffic profile composed by a large amount of low priority traffic. | Low | High | High |
| Resource _utilization_ with a traffic profile composed by a large amount of high priority traffic | Low | Low | High |
| TCs isolation. | High | Medium | Low |
| **BAM Operational Characteristics** | **MAM** | **RDM** | **ATCS** |
| "High-to-Low" (HTL) sharing. | No | Yes | Yes |
| "Low-to-High" (LTH) sharing. | No | No | Yes |

RDM and ATCS have the main objective of maximize resource (link, lambda, slot) utilization and, to achieve this goal, they allow resource sharing among traffic classes (TCs). This behavior reduces traffic isolation among traffic classes.

RDM model allows the sharing of non-allocated resources belonging to high priority traffic classes by low priority traffic classes (HTL sharing - Table 1) [4]. This model tends to improve resource utilization for a network traffic profile in which a large volume of low priority TCs and/or applications is demanding network resources.

AllocTC-Sharing (ATCS) model keeps RDM resource allocation strategy of "High-To-Low" loans and adds the possibility of "Low-To-High" loans (LTH sharing – Table 1). As such, AllocTC-Sharing allows high priority classes (TCs) to get resources normally used by low priority classes (TCs). In brief, "loans" are allowed in both directions (HTL e LTH). This model targets networks in which the input traffic has a highly dynamic profile with weak isolation among TCs being acceptable. This corresponds, as an example, to networks with high priority elastic applications like multimedia services, among others [4][5].

### IV. RESOURCE ALLOCATION FOR ELASTIC OPTICAL NETWORKS WITH BAMs (EON-BAM)

For the remaining of this paper, the proposed BAM-based resource allocation strategy is mentioned as EON-BAM.

EON-BAM approach requires a "BAM parameters mapping", Traffic Class (TC) definition and BAM model or "behavior" choices by the network operation and management for the specific EON scenario.

In terms of the BAM parameters mapping and for simplicity, the "manageable resource" defined for allocation and/or sharing in EON is the optical frequency slot [1][2][6][13]. It is noteworthy that, depending on the level of abstraction (scenario modeling defined), nothing prevents to consider other optical manageable resources in the BAM to EON-BAM mapping, such as optical converters and transceivers.

Traffic Classes (TCs) are defined and configured by the network operation and management instance and, considering the target of this paper, 03 TCs are defined as follows:

• Bronze, Silver and Gold Classes (equivalent to TC0, TC1 and TC2 in the simulation scenario following).

• TC0 (bronze) class aggregates applications, as an example of preliminary mapping, with low priority and possibly no or less demanding constraints.

• TC1 (silver) class aggregates applications with intermediate priorities and more restricted constraints in relation to TC0.

• TC2 (gold) class aggregates applications with higher priority and more restricted constraints.

To the extent of network resource allocation being evaluated, the specific applications mapped to each class may be left to the network operation and management instance. As already indicated in previous sections, applications within each specific class will compete for resource and may make use of resource available in other classes by BAM model arbitration.

The BAM model to be used ("behavior") is the third aspect to consider in the EON-BAM approach. Table II shows a set of possible EON-BAM behaviors mappings in relation to traffic profile. In other words, Table II indicates what EON-BAM slot utilization can be expected depending on which BAM was defined (configured) for slot allocation for distinct traffic profiles. It is relevant to observe that the expected traffic profile plays an important role since BAM´s behaviors are directly dependent of traffic. That also means EON-BAM must be evaluated (simulated) against the various BAM alternatives.

TABLE II. EON-BAM EXPECTED LINK UTILIZATION AND TRAFFIC PROFILE

| EON-BAM "Behavior" Mapping versus Traffic Profile | MAM | RDM | ATCS |
|---|---|---|---|
| *Slot utilization* with a traffic profile composed by a large amount of low priority traffic | Low | High | High |
| *Slot utilization* with a traffic profile composed by a large amount of high priority traffic | Low | High | High |
| TCs isolation | High | Low | Low |
| **EON-BAM Operational Characteristics** | **MAM** | **RDM** | **ATCS** |
| **Sharing of slots** "High-to-Low" (HTL) | No | Yes | Yes |
| **Sharing of slots** "Low-to-High" (LTH) | No | No | Yes |

Next section describes and discusses the simulation results evaluating the EON-BAM resource allocation with distinct BAMs for a network topology under different traffic patterns.

V. SIMULATION AND RESULTS

This section presents a case study evaluating the configuration of three different BAMs (MAM, RDM and ATCS) for EON-BAM in a partial NSFNet network topology. Such study aims to evaluate and compare the alternatives for EON-BAM in Elastic Optical Networks varying the BAM adopted and the input traffic pattern in four different simulation scenarios.

Figure 1a illustrates the NSFNet network topology. Figure 1b shows the nodes of the NSFNet topology used in our simulations considering node 14 as the source node of all optical connections requests (lightpaths) and nodes 2, 7 and 5 as destination nodes of lightpaths for Bronze, Silver and Gold traffic classes. The simulation topology (Figure 1b) and the traffic pattern modeled is intended to simplify the routing and the impact of continuity aware spectrum allocation that is out of scope in this preliminary evaluation work. It is important to remark that previous simulated BAM switching approaches (discussed in [12]) with IP networks effectively show their utilization characteristics with partial topologies. Since the objective is to evaluate the utilization improvement with distinct BAMs, it basically requires a scenario where the traffic pattern experiences moments of congestion for all classes.

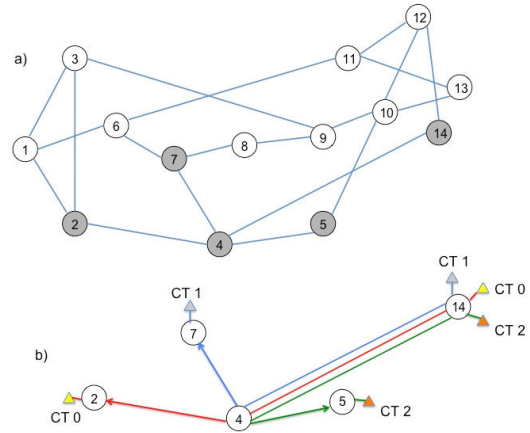

Fig. 1. EON-BAM Simulation Topology.

The configuration parameters of the simulation scenarios are as follows:

- Link: 400 slots (OFDM);
- Traffic Classes: TC0 (Bronze), TC1 (Silver) and TC2 (Gold);
- Resource Constraints, according to Tables III and IV.

The Traffic Classes have the following characteristics:

- TC0 (Bronze) – Path 14, 4, 2; 80 slots of maximum capacity (20% of network resources); requests of 10Gbps using BPSK modulation, which is equivalent to one slot.
- TC1 (Silver) – Path 14, 4, 7; 120 slots of maximum capacity (30% of network resources); requests of 40Gbps using BPSK modulation, which is equivalent to 2 slots.
- TC2 (Gold) – Path 14, 4, 5; 200 slots of maximum capacity (50% of network resources); requests of 100Gbps using BPSK modulation, which is equivalent to 5 slots.

TABLE III. RESOURCE CONSTRAINT PER TRAFFIC CLASS (TCS) - MAM

| RC | Max RC (%) | MAX RC (Slots) | TC |
|---|---|---|---|
| RC0 | 20 | 80 | TC0 (Bronze) |
| RC1 | 30 | 120 | TC1 (Silver) |
| RC2 | 50 | 200 | TC2 (Gold) |

TABLE IV. RESOURCE CONSTRAINT PER TRAFFIC CLASS (TCS) - RDM

| RC | Max RC (%) | MAX RC (Slots) | TC |
|---|---|---|---|
| RC0 | 100 | 400 | TC0 (Bronze) |
| RC1 | 80 | 320 | TC1 (Silver) |
| RC2 | 50 | 200 | TC2 (Gold) |

In all evaluated scenarios, the hold time of lightpaths is modeled exponentially with a mean of 2500h. The spectrum allocation algorithm used was First-Fit [14]. There were generated one million lightpaths requests with 10 replications.

The scenarios evaluated are as follows:

- Scenario 01: arrival rate for TCs of higher priority is higher than arrival rate for TCs of lower priority. Traffic

generated is initially higher for TCs of lower priority and then gets higher for all classes.

- Scenario 02: arrival rate for TCs of higher priority is higher than arrival rate for TCs of lower priority. Traffic generated is initially higher for TCs of higher priority and then gets higher for all classes.

- Scenario 03: arrival rate for TCs of lower priority is higher than arrival rate for TCs of higher priority. Traffic generated is initially higher for TCs of lower priority and then gets higher for all classes.

- Scenario 04: arrival rate for TCs of lower priority is higher than arrival rate for TCs of higher priority. Traffic generated is initially higher for TCs of higher priority and then gets higher for all classes.

In all scenarios, the following performance metrics were evaluated:

- Blocking – computed unestablished lightpaths;
- Slots utilization – the effective slot utilization; and
- Established lightpaths – number of established lightpaths.

### A. Scenario 01 - Description and Results Evaluation

In this scenario, the simulation parameters are as follows:

- lightpath requests – TC0 (Bronze): generated every 40h with a delay of 5000h at the first request. In other words, after 5000h from the first request, it starts TC0 requests;
- lightpath requests – TC1 (Silver): generated every 20h with a delay of 3000h at the first request; and
- lightpath requests – TC2 (Gold): generated every 10h, without delay at the first request.

Figures 2 and 3 show the number of blocked requests and the number of established requests by configured BAM model.

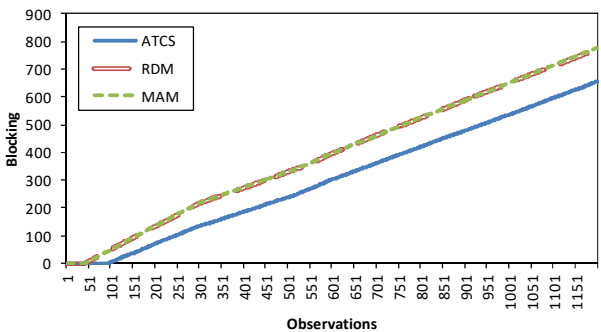

Fig. 2. Blocked Requests – Scenario 01.

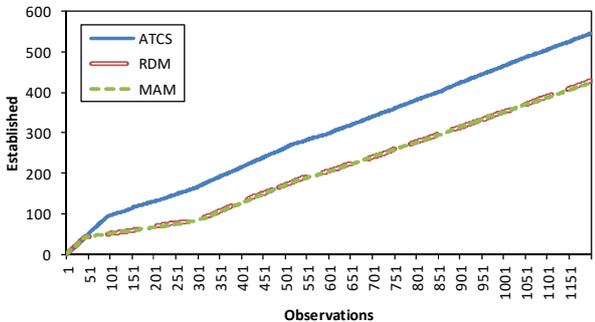

Fig. 3. Established Requests – Scenario 01.

It can be noticed in Fig. 2 that the ATCS BAM model has lower number of blocked lightpaths when compared to other BAM models. This behavior is explained by the fact that in this scenario are generated more requests of high priority classes. The ATCS model allows better resource sharing among classes through loans, such as "low-to-high", that does not occur with other BAM models. Consequently, as illustrated in Fig. 3, the ATCS model succeeds to establishes a greater number of lightpath requests.

In Figure 4, MAM and RDM models show similar behavior in terms of slot utilization. This is because the traffic class with highest priority was generated first with lowest arrival times. In this scenario, the ATCS model presents greater slots utilization by allowing "low-to-high" loans.

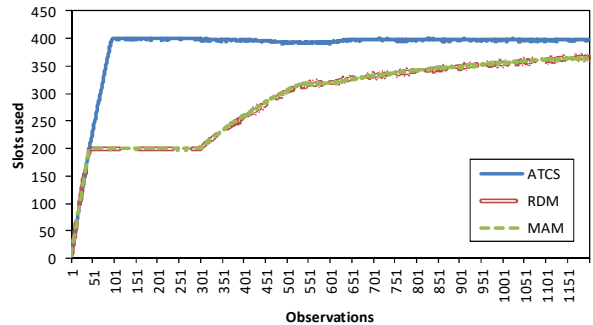

Fig. 4. Slots Utilization – Scenario 01.

### B. Scenario 02 - Description and Results Evaluation

In this scenario, the simulation parameters are as follows:

- *lightpath requests* – TC0 (Bronze): generated every 40h, without delay at the first request;
- *lightpath requests* – TC1 (Silver): generated every 20h with a delay of 3000h at the first request;
- *lightpath requests* – TC2 (Gold): generated every 10h, with a delay of 5000h at the first request.

Figures 5 and 6 show the number of blocked requests and the total number of requests for each configured BAM.

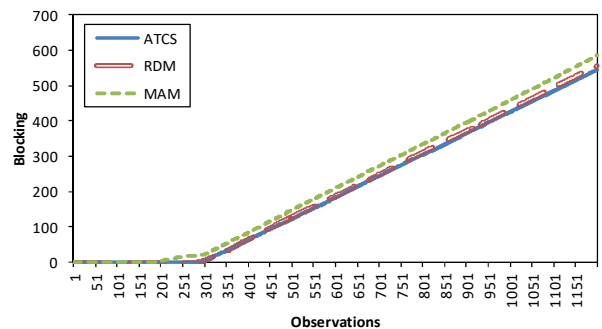

Fig. 5. Blocked Requests – Scenario 02.

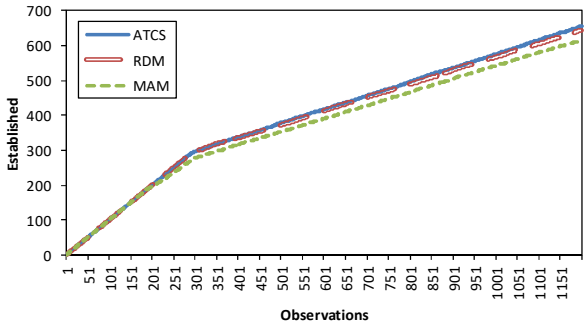

Fig. 6. Established Requests – Scenario 02.

It is noted in Fig. 5 that the BAM models have similar values for blocked lightpaths. Such behavior is justified mainly by the fact that in this scenario, despite the higher rate of arrival of higher priority classes, these classes have a delay at the beginning of the simulated traffic. Consequently, as illustrated in Fig. 6, the resource allocation models attend a similar number of lightpaths requests.

In Figure 7 the evaluated models show a similar behavior in the use of slots. This is because the highest priority classes of requests are being generated delayed, despite the inter-arrival times are lower.

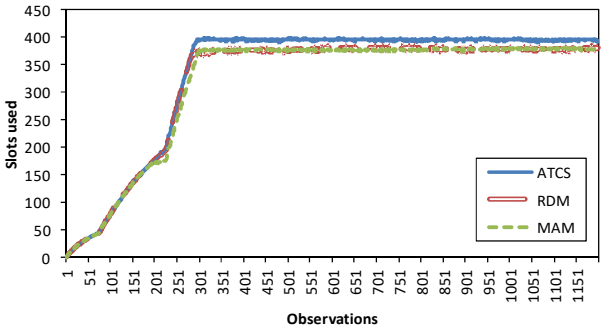

Fig. 7. Slots utilizations – Scenario 02.

### C. Scenario 03 - Description and Results Evaluation

In this scenario, the simulation parameters are as follows:

- *lightpath requests* – TC0 (Bronze): generated every 10h, with a delay of 5000h at the first request;
- *lightpath requests* – TC1 (Silver): generated every 20h with a delay of 3000h at the first request;
- *lightpath requests* – TC2 (Gold): generated every 40h, without delay at the first request.

The Figures 8 and 9 show the number of blocked requests and total requests for each configured BAM.

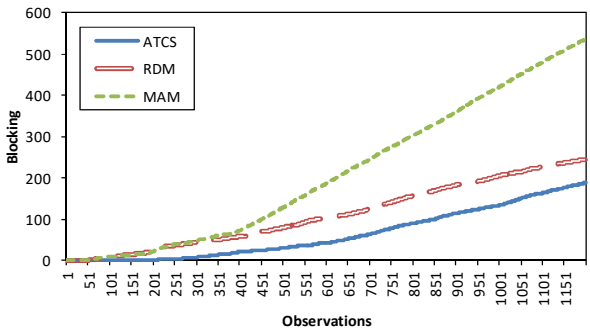

Fig. 8. Blocked Requests – Scenario 03.

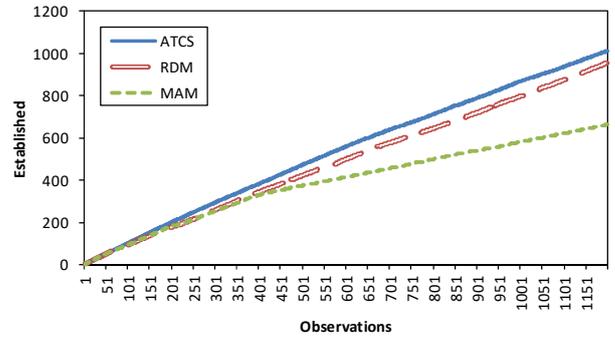

Fig. 9. Established Requests – Scenario 03.

It is noted in Fig. 8 that the ATCS model presents lower lightpaths blocking when compared with other bandwidth allocation models. Such behavior is justified by the fact that in this scenario are first generated requests for higher priority classes and in this scenario, the ATCS model allows better sharing of resources among the classes, through "low to high" loans, which does not occur with other evaluated models. Consequently, as illustrated in Fig. 9, with the ATCS model it can be established a greater number of lightpaths requests.

In Figure 10 is possible to observe that the evaluated BAM models show a similar behavior in term of slots utilization. This is because the highest priority classes of requests are being first generated but with longer inter-arrivals time. In this scenario, the ATCS model presents at the beginning of observation, the greater use of slots allowed by "low to high" loans.

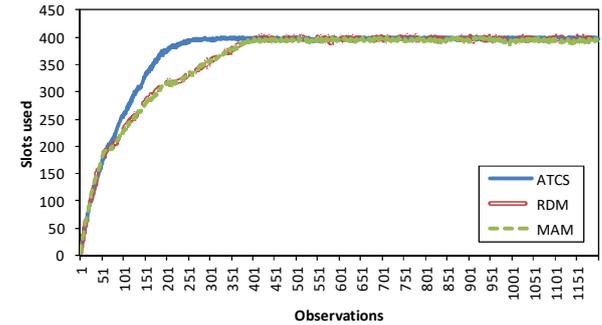

Fig. 10. Slots utilizations – Scenario 03.

### D. Scenario 04 - Description and Results Evaluation

In this scenario, the simulation parameters are as follows:

- *lightpath requests* – TC0 (Bronze): generated every 10h, without delay at the first request;
- *lightpath requests* – TC1 (Silver): generated every 20h with a delay of 3000h at the first request;
- *lightpath requests* – TC2 (Gold): generated every 40h, with a delay of 5000h at the first request.

It is noted in Fig. 11 that the MAM model showed higher lightpath blocking when compared with other BAMs. Such behavior is justified because in this scenario are generated more lower priority classes requests. In this scenario, the MAM model does not allow the sharing of resources among classes, which does not occur with other configured models. Both RDM and ATCS allow the "top down" sharing resulting in lower request blocking. Consequently, as illustrated in Fig. 12, the MAM model establishes a smaller number of lightpath requests when compared with other evaluated BAM models.

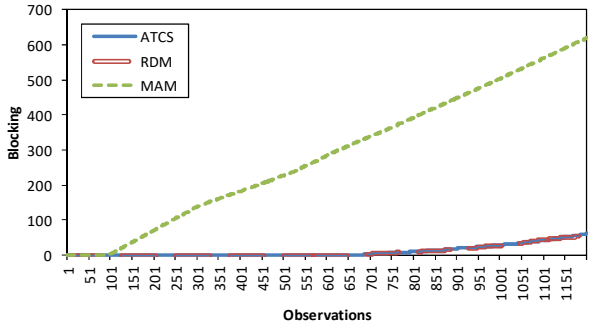

Fig. 11. Blocked Requests – Scenario 04.

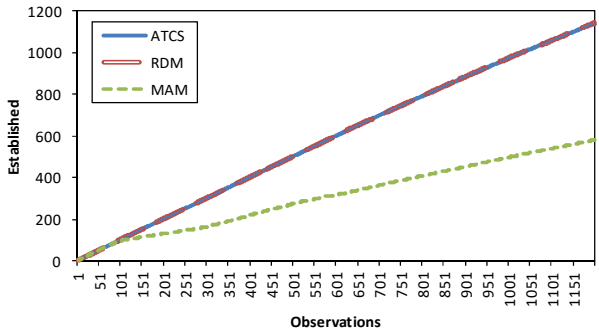

Fig. 12. Established Requests – Scenario 04.

Figure 13 shows slot utilization per BAM. It is possible to observe that the ATCS and RDM models show a similar behavior in terms of slot utilization. This occurs because the highest priority class requests are being generated at last with higher inter arrival times. In this scenario, the ATCS and RDM models present increasing slot utilization after simulation's initial phase by "top down" loans enabled, which do not occur with the MAM model.

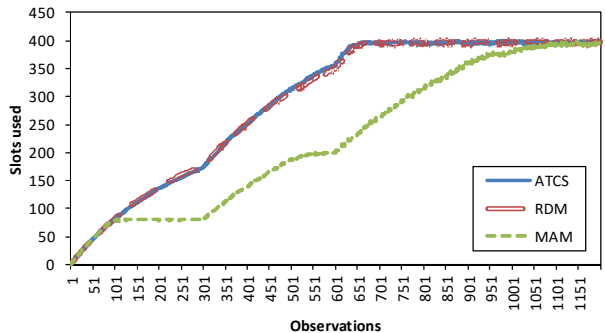

Fig. 13. Slots utilization – Scenario 04.

## VI. Final Consideration

This paper presented a preliminary evaluation of slot allocation based on Bandwidth Allocation Models for Elastic Optical Networks (EON-BAM).

Three BAMs (MAM, RDM and AllocTC-Sharing) were configured and simulated for slot allocation with distinct traffic scenarios. The traffic scenarios simulated tried to reflect the variability of input traffic configuration currently existing on EON.

In general, the consistency of results in different traffic scenarios indicates the potential applicability, flexibility and, consequently, better utilization of EON's resource when EON-BAM is used.

When we consider the alternatives among the evaluated BAM models for EON-BAM, the AllocTC-Sharing "behavior" shows the best possible result. It is observed that this model deals with all traffic profiles and "adapts" itself on a dynamic and opportunistic way, sharing slots among high priority and low priority traffic classes. AllocTC-Sharing configuration for EON-BAM leads to a better resource utilization when compared with other BAM models.

It is also important to mention that this is a preliminary result and conclusion is focused on resource utilization only. In effect, in case AllocTC-Sharing is configured for EON-BAM, managers and network polices must consider the fact that "slot devolutions" will occur. As such, the application and services mapping to the configured traffic classes (TCs) must consider the fact that devolutions will eventually occur and applications mapped to high priority classes should support an "elastic" behavior.

Another open issue for future work is that link utilization has a better performance with AlocTC-Sharing but this BAM model needs to be evaluated against other non-BAM-based algorithms and solutions available in the literature.